\documentclass{eptcs}
\providecommand{\event}{TFPIE 2018} % Name of the event you are submitting to
\usepackage{breakurl}             % Not needed if you use pdflatex only.
\usepackage[noxcolor]{pstricks}
\usepackage{pst-node}
\usepackage{graphicx}
\usepackage{subfigure}
\usepackage{stfloats}
\usepackage{multido}
\usepackage{setspace}
\usepackage{textcomp}
\usepackage{amssymb}
\usepackage{alltt}

\title{Vector Programming Using Generative Recursion \\ \Large{A Framework for Beginners Using Vector Intervals}}

\author{Marco T. Moraz\'an
\institute{Seton Hall University}
\email{morazanm@shu.edu}}
\def\titlerunning{Vector Programming Using Generative Recursion}
\def\authorrunning{M. T. Moraz\'an}

\begin{document}
\maketitle

\begin{abstract}
Vector programming is an important topic in many Introduction to Computer Science courses. Despite the importance of vectors, learning vector programming is a source of frustration for  many students. Much of the frustration is rooted in discovering the source of bugs that are manifested as out-of-bounds indexing. The problem is that such bugs are, sometimes, rooted in incorrectly computing an index. Other times, however, these errors are rooted in mistaken reasoning about how to correctly process a vector. Unfortunately, either way, all too often beginners are left adrift to resolve indexing errors on their own. This article extends the work done on vector programming using vector intervals and structural recursion to using generative recursion. As for problems solved using structural recursion, vector intervals provide beginners with a useful framework for designing code that properly indexes vectors. This article presents the methodology and concrete examples that others may use to build their own CS1 modules involving vector programming using any programming language.
\end{abstract}

\maketitle

\section{Introduction}
Undoubtedly, many readers of this article remember enduring endless hours debugging programs that manipulate vectors. Many of those hours were probably spent trying to determine why an index into a vector was out of range. The same readers probably also remember how irritating it was to discover that the bug was trivial or that most of the time was spent searching for the root of the bug in the wrong function.  Surprisingly, this scenario still occurs all too often among beginners. At the heart of the problem is that beginners are not provided with a framework that they can use to design and debug vector-processing (a.k.a. array-processing) functions.

The vector-programming frustration beginners endure, of course, affects the quality of their design and of their code. This makes it difficult for others to understand how they solved a problem--a fundamental pillar of programming \cite{PPL}. Furthermore, they develop the bad habit to patch their code instead of properly designing their code. Given the importance of vectors in Computer Science this is worrisome. Vectors allow us to efficiently implement algorithms for applications in a myriad of fields. Vectors, for example, allow us to efficiently implement various data structures such as binary trees \cite{Ford}, stacks \cite{Goodrich}, and process queues \cite{Silberschatz}. They also allow us to reduce the complexity of algorithms such as, for example, finding a path in a directed graph \cite{HtDP} and matrix multiplication \cite{Burden}. Vectors are also useful in the reduction of memory allocation by, for example, allowing us to sort, say, files or numbers in place \cite{Knuth3}. Given the importance and versatility of vectors it is in our interest as a community to make an introduction to vector programming for beginners as frustrationless as possible and to provide an environment in which students learn solid design techniques to write code. This can be achieved by providing beginners with a structured model they can use to reason about processing vectors.

%Much of the introductory material to vector programming suffers from offering a structureless definition for vector programming and from focusing mostly on the syntax for using vectors in a given programming language. Some material goes a step further and presents an \textsf{ADT} for vectors. Some of these \textsf{ADT}s, however, describe a single operation: indexing. It is, thus, not surprising that many beginners feel they are left adrift to figure out on their own how to avoid out-of-bounds indexing errors. This is highly undesirable for at least two reasons. The first is that students get frustrated enough to quit Computer Science as a major. This is an especially important issue in universities, like in the USA, where beginning students spend a semester or two shopping around for a major. The second is that students that persevere  develop bad programming habits associated with the belief that vector programming is a strict exercise in trial-and-error instead of an exercise in design. In the interest of absolute clarity, trial-and-error has its place in programming. A programmer ought to experiment with different designs by trial-and-error, but should not write code--with no design--by trial-and-error. The emphasis ought to be on design and not on hacking code.

This article builds on the type-driven work developed at Seton Hall University (\textsf{SHU}) to introduce students to vector programming using vector intervals and structural recursion \cite{VINTV}. In a type-driven approach, the type of data being processed guides the design of a solution. At the heart of the approach presented in this article is providing students with data definitions that help them design vector-processing functions to solve problems. In a program, vectors are declared with an arbitrary fixed size. Therefore, these data definitions must be recursive data definitions. Such data definitions can be directly exploited using structural recursion to develop programs. As this article demonstrates, the same framework is used to design and develop vector-processing programs using generative recursion. The framework taught to beginning students is based on the concept of a \emph{vector interval} that may only contain valid indices into a vector. At the heart of the approach is to have students demonstrate that an index into a vector is contained in a valid vector interval and that any new vector interval generated (e.g., for a recursive call) is valid for a given vector. The article is organized as follows. Section \ref{rw} discusses related work. Section \ref{sb} discusses the programming background of the students with whom this approach is used. Section \ref{srec} describes and illustrates the use of vector intervals using structural recursion. Section \ref{examples} discusses vector processing issues that arise when using generative recursion by presenting the in-class development of three extended vector-programming examples. These example are focused around sorting a vector in-place using algorithms that some may believe are too difficult for beginners. Finally, Section \ref{conclusions} presents concluding remarks and directions for future work.

\section{Related Work}
\label{rw}
Many textbooks introduce vectors as lacking a recursive structure that can be exploited to solve problems. Readers are then introduced to programming with vectors using snippets of code that emphasize that an index into the vector must be within its bounds. For example, a vector is described as a collection of variables of the same type with each element having an index \cite{Goodrich} or as a finite sequential list of elements of the same datatype identifying the first element, the second element, the third element, and so forth \cite{Ford}. Such data definitions are inadequate, because they hide the recursive nature of the interval of valid indices into the vector and focus exclusively on the syntax to declare, create, index, and mutate vectors. They fail to provide the proper model to help beginners design vector-processing functions/methods that avoid, for example, illegal indexes into the vector. The second data definition may even be considered misleading by describing a vector as a list that has a well-known recursive structure. Vectors do not have a decomposable structure like lists. Furthermore, describing vector elements as the first, the second, the third, and so forth does not assist in any way the design of vector-processing functions. We can not program ``and so forth." Even some modern approaches to make programming popular among the young address vectors in a very similar manner (e.g., \cite{codeorg}). In contrast, the work presented in this article aims to provide students with a recursive decomposable data definition that they can use to reason about vector processing when using generative recursion. This data definition is that of a vector interval \cite{VINTV}. A vector interval only contains valid indices into a given vector and, thus, is a useful abstraction to determine that a vector is properly indexed.

The problem of only using legal indices into a vector is traditionally and summarily left to the student with no clear indication of how to accomplish this task (e.g., \cite{Sedgewick}). Some may convincingly argue that this is relatively easy when you need to process an entire vector. However, the matter is not clear when you need to process only part of a vector. It is even less clear when vectors are processed using generative recursion. Consider sorting a vector using quicksort which requires partitioning and independently sorting different parts of the vector. It is difficult for a beginner to determine or to be confident that she is always correctly indexing the vector. Even worse, it is more difficult to pin down bugs when indexing errors occur if a model that helps the student reason about indexes is not provided to them. Vector intervals provide students such a model. If the vector interval is empty then students know that the vector should not be indexed. If the vector interval is not empty, students know that they must only use an index that they can demonstrate is in the given vector interval. Therefore, out-of-bounds errors must be the result of giving a function an improper/invalid vector interval to process. This is one of the most powerful features of designing vector-processing functions using vector intervals.

Some efforts have gone beyond syntax. The textbook \emph{How to Design Programs} (\textsf{HtDP}), for example, describes a vector as a well-defined mathematical class of data with some basic constructors and observers \cite{HtDP}. \textsf{HtDP} further states that we can think of vectors as functions on a small finite range of natural numbers. This begins to provide some context for vector-processing, but surprisingly falls short of identifying the recursive structure of this range of natural numbers as it does so well for other types of data. It is unlikely that future editions of \textsf{HtDP} are to develop this given that the second edition has eliminated its introduction to vector programming \cite{HtDP2}. In contrast, the work presented in this article tackles the recursive nature of this finite range of natural numbers and describes it as an interval. Furthermore, the fact that this range is finite is carried to its logical conclusion to obtain two data definitions for a vector interval. One data definition leads to a function template that is used to design functions to process vectors from right to left (i.e., from the largest index down to the lowest index in the vector interval) and the other is used to design functions that process vectors from left to right (i.e., from the lowest index to the largest index in the vector interval). It is, of course, possible to design other access patterns into a vector and these make excellent exercises for students to practice their design skills.

Related to the material discussed in this article, although not directly, are ubiquitous efforts to hide low-level vector manipulation from programmers. For example, an \textsf{Iterator} generates a sequence of elements from a collection, one at a time \cite{Tymann}. This allows programmers to examine every element of a vector without worrying about properly indexing the vector. The details of properly indexing the vector are hidden by the \textsf{Iterator}. Similarly in functional programming, for example in \textsf{Racket}, programmers have access to functions like \textsf{vector-map} and \textsf{vector-filter} that iterate through a vector and that hide the details of properly indexing the vector. Clearly, iterators and higher-order vector-processing functions process the entire interval of valid indices into a vector, but are not useful when a proper subinterval must be processed. For example, processing proper subintervals are common in divide-and-conquer sorting and searching algorithms (e.g., quick and heap sorting and binary search). In contrast, programming using vector intervals allows programmers to reason about and write programs that process proper subintervals of all valid indices. In fact, the entire interval of valid indices into a vector is an instance of a vector interval. Thus, vector intervals can be used as the foundation to program the abstractions provided by iterators and higher-order vector-processing functions.

Vector intervals were first defined to design vector processing functions using structural recursion \cite{VINTV}. They are defined in terms of the size, \textsf{N}, of a vector. The idea is that a valid interval of indices into a vector must be in [0..N-1]. This notion is similar to that found in dependently-typed programming languages such as Dependent ML \cite{Xi} and Idris \cite{Brady}. Vector intervals as used in this article (in the context of the \textsf{HtDP} student languages) do not enforce type invariants as done by a type-based compiler, but do plant in a student's mind that data definitions can depend on run-time properties of the data processed and force students to engage in type-oriented design.

\begin{figure*}[t]
\begin{alltt}
     ; f-on-vector: (vector X) \(\rightarrow\) \(\ldots\)
     ; Purpose: \(\ldots\)
     (define (f-on-vector V)
         (local [; f-on-VINTV: VINTV(wholenum wholenum) \(\rightarrow\) \(\ldots\)
                 ; Purpose: For the given VINTV, \(\ldots\)
                 (define (f-on-VINTV low high)
                   (cond [(empty-VINTV? low high) \(\ldots\)]
                         [else \(\ldots\) (vector-ref V high) \(\ldots\)
                               \(\ldots\) (f-on-VINTV low (sub1 high))]))

                 ; f-on-VINTV2: VINTV2(wholenum wholenum) \(\rightarrow\) \(\ldots\)
                 ; Purpose: For the given VINTV2, \(\ldots\)
                 (define (f-on-VINTV2 low high)
                   (cond [(empty-VINTV2? low high) \(\ldots\)]
                         [else \(\ldots\) (vector-ref V low) \(\ldots\)
                               \(\ldots\) (f-on-VINTV2 (add1 low) high)]))]
           \(\ldots\)))
\end{alltt}
\caption{The Template for Functions on Vectors.}
\label{vtemplate}
\end{figure*}

\section{Student Background}
\label{sb}
At \textsf{SHU}, the Introduction to Computer Science spans two semesters and focuses on problem solving using a computer \cite{mtm22,mtm24}. The languages of instruction are the successively richer subsets of \textsf{Racket} known as the student languages which are tightly-coupled with \textsf{HtDP} \cite{HtDP,HtDP2}. No prior programming experience is assumed. Before introducing students to vector programming, the first course familiarizes students with primitive data (e.g., numbers, strings, booleans, symbols, and images), primitive functions, and library functions to manipulate images (i.e., the image teachpack). During this introduction, students are taught about variables, defining their own functions, and the importance of writing contracts and purpose statements. The next step of the course introduces students to data analysis and programming with compound data of finite size (i.e., structures). At this point, students are introduced to the first design recipe. Students gain experience in developing data definitions, examples for data definitions, function templates, and tests for all the functions they write. A great deal of emphasis is placed on all of these steps as part of the problem-solving design process. Building on this experience, students develop expertise on processing compound data of arbitrary size such as lists, natural numbers, and trees. In this part of the course, students learn to design functions using structural recursion. After structural recursion, students are introduced to functional abstraction and the use of higher-order functions such and \textsf{map} and \textsf{filter}. The first course ends with a module on distributed programming \cite{mtm26,mtm27}.

In the second course students are exposed to generative recursion, accumulative recursion, and mutation \cite{mtm24,mtm25}. The course starts with generative recursion. At the end of this module, students get their first exposure to vector programming. Students are taught the syntax needed for vectors and are introduced to the design of vector-processing functions using vector intervals and structural recursion.  After this, the course exposes students to accumulative recursion and iteration. The course ends with two modules on mutation that include their second exposure to vector programming. In this second exposure, students design vector mutators using vector intervals and both generative and accumulative recursion.

The topics covered follow much of the structure of \textsf{HtDP} \cite{HtDP2}. There are two 75-minute lectures every week and the typical classroom has between 20 to 25 students. In addition to the lectures, the instructor is available to students during office hours (at least 3 hours/week) and there are tutors available 20-30 hours each week. Students voluntarily visit the tutors as much as they need. The tutoring hours are conducted by undergraduate students handpicked and trained by the author. These tutors focus on making sure students develop answers for each step of the design recipe (from writing contracts to running tests). Students must attempt to follow the steps of the design recipe prior to attending tutoring. Based on a student's work, the tutors provide guidance but do not solve problems. Students are still responsible for successfully completing all steps of the design recipe. In addition, tutors attend lectures to assist students when they get stuck with, for example, syntax errors. This type of team-teaching with undergraduate tutors has proven to be extremely well-received by students and to be an effective means to enhance the learning experience.

\begin{figure*}[t]
\begin{alltt}
   ; avg-vector: (vectorof number) \(\rightarrow\) number
   ; Purpose: To compute the average of the given vector
   ; Assumption: The vector is not empty.
   (define (avg-vector V)
     (local [; sum-elems: VINTV(wholenum wholenum) \(\rightarrow\) natnum
             ; Purpose: For the given interval, sum the vector elements
             (define (sum-elems low high)
               (cond [(empty-VINTV? low high) 0]
                     [else (+ (vector-ref V high)
                           (sum-elems low (sub1 high)))]))]
       (/ (sum-elems 0 (sub1 (vector-length V))) (vector-length V))))

   (check-within (avg-vector (vector 6 7 8 9)) 7.5 0.01)
   (check-within (avg-vector (vector 1 2 3)) 2 0.01)
\end{alltt}
\caption{A Function to Compute the Average of a Vector of Numbers.}
\label{avgv}
\end{figure*}

\section{Vector Processing Using Structural Recursion}
\label{srec}
For a given vector \textsf{V} of size \textsf{N}, we may want to process the entire vector (indices in [0..\textsf{(sub1 N)]} or we may want to process only part of the vector (indices in [\textsf{a}..\textsf{b}]). Clearly, the vector indices that must be processed define an interval. We must design functions, however, in a manner that avoids accessing a vector with an index outside the given interval. This requires developing a data definition for a  \emph{vector interval}. A vector interval is an interval that places restrictions on what values the indices \textsf{low} and \textsf{high} of the vector interval may take. In general, a valid index into a vector, \textsf{V}, of size \textsf{N} is between 0 and \textsf{(sub1 N)}. Thus, we can define a vector interval as follows:
\begin{alltt}

  For a vector of size N, a vector interval, VINTV(low..high), is two
  whole numbers, low \(\geq\) 0 and -1 \(\leq\) high \(\leq\) N-1, such that it is  either:

       1. empty  (i.e., low > high)
       2. VINTV(low..high-1)..high

\end{alltt}
This is a recursive structural data definition. It states that a vector interval, VINTV(low high), may be empty (base case) or it can be non-empty (recursive case). In the non-empty case, we have the \textsf{high} element and the sub-vector-interval, VINTV(low..high-1), that contains all elements but \textsf{high}. Observe that there is no need to specify an upper bound for \textsf{low}, like \textsf{low $<$ N}, because if \textsf{low} ever failed to satisfy this condition the vector interval would be empty.

Using this data definition, to process some vector \textsf{V}, means that in the body of a function to process the \textsf{VINTV}, \textsf{V} must be referenced. Any index into \textsf{V} must be a member of the \textsf{VINTV}\footnote{It may also be a member of a containing valid vector interval when using generative recursion.}. Furthermore, it means that the vector interval is processed from right to left. A similar development takes place for a data definition, \textsf{VINTV2}, to process vector elements from left to right (i.e. from low to high) \cite{VINTV}.

The above observations allow for the in-class development of the function template to process a vector displayed in Figure \ref{vtemplate}. The contract states that any function that processes a vector must take as input at least a vector of any type (\textsf{X} is a type variable). The body of the function is a \textsf{local}-expression that may be used to define one or more local functions and values. Students are told that problem analysis will reveal the type of expression that is needed in the body of the \textsf{local}-expression. If a single value is needed from the given vector, then the expression will be one that processes a vector interval. Otherwise, the expression will be one that uses different values obtained from processing the same vector. The local definition section contains two templates: one for each direction that a vector interval can be processed in. At least one of the templates is to be used to process vector elements.

To make the use of the function template concrete, consider computing the average of a vector of numbers. Problem analysis reveals that the vector cannot be empty given that division by 0 is undefined. Now, the template for functions on a vector from Figure \ref{vtemplate} is specialized. The contract indicates that the input is a vector of numbers, \textsf{V}, and that the function returns a number. The body of the \textsf{local}-expression must be an expression that divides the sum of the vector elements by the length of the vector. This means that we must write a function to compute the sum of vector elements. Given our problem analysis, either of the templates to process a vector interval can be used. It does not matter in which direction the vector interval is processed as addition is a commutative operation. Without loss of generality, we can choose to process from right to left (i.e., the template for \textsf{VINTV}) in the auxiliary function \textsf{sum-elems}. This means that when the vector interval is empty the answer is 0 and that when it is not empty we add \textsf{(vector-ref V high)} to the result of recursively processing the rest of the \textsf{VINT} (i.e., \textsf{VINTV(low..high-1)}). The resulting function is displayed in Figure \ref{avgv}. Observe that by using the template based on structural recursion it is impossible to have indexing errors if a valid vector interval is provided as the initial input to \textsf{sum-elems}. Thus, simplifying the job of designing vector processing functions for beginners. The initial vector interval, of course, needs to be the entire interval of valid indexes as shown in the body of the \textsf{local} in Figure \ref{avgv}.

\section{Vector Processing Using Generative Recursion}
\label{examples}
Many efficient solutions to problems are not based on structural recursion. Instead, they are based on generating new instances of the problem (which are, hopefully, easier to solve). This adds complexity to vector programming as follows:
\begin{enumerate}
  \item Students must make sure that any new vector interval generated is valid.
  \item Students must make sure that any index into a vector is a member of a valid vector interval.
\end{enumerate}
In this context, indexing errors may arise by either providing an invalid initial vector interval, generating an invalid vector interval, or by incorrectly indexing the vector with a value outside a valid vector interval. The second and third concerns are, of course, new to students as they can never arise when correctly using structural recursion. The key to success, therefore, is to prove that any vector interval (initial or generated) is valid for a given vector and to prove that an index into the vector is in the interval [\textsf{low} to \textsf{high}].

Designing vector-processing functions using the concept of a vector interval, however, proves useful in avoiding indexing errors when using generative recursion.  To illustrate how this is done in class, three extended examples to sort a vector in-place are presented: quicksort, heapsort, and radixsort.

\subsection{Quicksort}
Students have been previously exposed to mutation-free quicksort using lists. Sorting a vector in place requires careful problem analysis in class. After some discussion, the algorithm to quicksort a vector of numbers in-place is summarized as follows:
\begin{alltt}
     Pick a pivot (say, V[low])
     Partition the vector
          VINTV(low..i) contains the elements \(\leq\) pivot
          VINTV(i+1..high) contains the elements > vector
     Swap V[low] and V[i]
     Recursively sort VINTV(low..i-1) and VINTV(i+1..high)
\end{alltt}

\begin{figure*}[t]
\begin{alltt}
; qs-in-place\(!\): (vectorof number) \(\rightarrow\)  (void)
; Purpose: To sort the array in non-decreasing order.
; Effect: The elements of the array are rearranged in place.
(define (qs-in-place\(!\) V)
  (local [; partition\(!\): VINTV2(wholenum wholenum) natnum --> number
          ; Purpose: For the given VINTV, partition E and place pivot in
          ;          final position.
          ; Effect: Mutate V so that all elements before the pivot are
          ;         <= pivot and all elements after the pivot are > pivot.
          ; Assumption: The given VINTV is not empty
          (define (partition\(!\) low high pp)  ...)
          ; qs-aux\(!\): VINTV2(wholenum wholenum) \(\rightarrow\) (void)
          ; Purpose: For the given VINTV, sort V in non-decreasing order.
          ; Effect: The elements in the given interval are rearranged in
          ;         place.
          (define (qs-aux\(!\) low high)
            (cond [(empty-VINTV2? low high) (void)]
                  [else (local [(define pp (partition\(!\) low high low))]
                          (begin (qs-aux\(!\) low (sub1 pp))
                                 (qs-aux\(!\) (add1 pp) high)))]))]))
    (qs-aux\(!\) 0 (sub1 (vector-length V)))))
\end{alltt}
\caption{The Beginning Template Specialization for Quicksort.}
\label{qs}
\end{figure*}

The beginning of the template specialization for quicksort is displayed in Figure \ref{qs}. Students realize that there is only one task that needs to be done with the vector and, therefore, the body of the local only calls a function, \textsf{qs-aux!}, that processes the entire vector. It is provided with the valid vector interval \textsf{[0..(sub1 (vector-length V))]}. Given their experience with vector intervals using structural recursion, students know that if the vector interval is empty then the vector should not be indexed and the process is done returning \textsf{(void)} as required by the contract. If the vector interval is not empty, then it must be partitioned and two new vector intervals to be sorted must be generated. Partitioning is determined to be a different problem from sorting and, therefore, students quickly realize that it requires an auxiliary function which is separately designed. The remaining design question that must be resolved for \textsf{qs-aux!} is which new vector intervals to generate. Given our previous problem analysis, students quickly converge on \textsf{[low..(sub1 pp)]} and \textsf{[(add1 pp)..high]}, where \textsf{pp} is the position of the pivot in the sorted vector and is returned by the partitioning function. They are now asked if these intervals are valid. Students conclude that if \textsf{partition!} returns an index in the given vector interval, then the two generated vector intervals are valid. Therefore, an indexing error cannot occur and these vector intervals can safely be recursively processed. It is noteworthy to highlight that students are now confident that if indexing errors arise it must be related to the yet undesigned \textsf{partition!} mutator. That is, students know where they must look for the resolution of a bug. The reader can contrast this with other approaches that leave students adrift trying to determine where an indexing bug needs to be resolved.

\begin{figure*}[t]
\begin{alltt}
(define (partition\(!\) low high pivot-pos)
   (local
         [(define (swap i j) ...)
          (define (small-index low high pivot) ...)
          (define (large-index low high pivot) ...)
          ; separate\(!\): VINTV(wholenum wholenum) number \(\rightarrow\) natnum
          ; Purpose: For the given VINTV, separate smaller and larger elements
          ; Effect: Mutate V so that all elements before the pivot are
          ;         <= pivot and all elements after the pivot are > pivot.
          ; Assumption: The given VINTV is not empty
          (define (separate\(!\) low high pp)
            (local [(define s-index (small-index low high (vector-ref V pp)))
                    (define l-index (large-index low high (vector-ref V pp)))]
              (cond [(<= s-index l-index) s-index]
                    [else  (begin   (swap s-index l-index)
                                    (separate\(!\) l-index s-index pp))])))
     (begin
         (local [(define pp (separate\(!\) low high pivot-pos))]
            (begin (swap pivot-pos pp) pp)))))
\end{alltt}
\caption{The outline for \textsf{partition!}.}
\label{partition}
\end{figure*}

Problem analysis reveals that \textsf{partition!} must take as input a non-empty vector interval that must be partitioned and the position of the pivot. It is clear to students that the given vector interval must be a sub-interval, \textsf{low} and \textsf{high}, within the vector interval provided to \textsf{qs-aux!}. In addition, \textsf{partition!} must also take as input the position of the pivot which is arbitrarily chosen to be \textsf{low}. This design is reflected in Figure \ref{qs}. The beginning outline for the function \textsf{partition!} is displayed in Figure \ref{partition}. After some class discussion, students conclude that they need a mutator, \textsf{separate!}, that places in \textsf{V} the elements less than or equal to the pivot before the elements greater than the pivot and that returns the largest index, \textsf{pp}, that has an element less than or equal to the pivot. This index is the correct position of the pivot in the sorted array. The body of \textsf{partition!}, therefore, must only swap the \textsf{pp} and the pivot-position elements in \textsf{V}. Observe that if \textsf{separate!} returns an element of the vector interval given to \textsf{partition!}, then an indexing error can not occur. That is, both indexes provided to \textsf{swap} are valid.

\begin{figure}[t]
\begin{alltt}
            ; heap-sort\(!\): (vectorof number) \(\rightarrow\) (void)
            ; Purpose: To sort in non-decreasing order the given
            ;          vector in place using heap sort
            ; Effect: Vector elements reorganized to be sorted
            (define (heap-sort\(!\) V)
              (local [\(\ldots\)]
                (begin
                  (heapify\(!\) \(\ldots\) \(\ldots\))
                  (sort\(!\) 0 (sub1 (vector-length V))))))
\end{alltt}
\caption{The outline for heapsort.}
\label{hs}
\end{figure}

The mutator \textsf{separate!} takes as input the vector interval provided to \textsf{partition!} and the index to the pivot. Observe that the vector interval provided is valid and not empty. In-class problem analysis reveals that smaller vector subintervals, within the given vector interval, need to be processed until the subinterval has a length less than 2. To achieve this the vector interval provided needs to be processed twice at each step: once from the right and once from the left. From the right we are searching for, if it exists, the largest index to a value less than or equal to the pivot. If such an index does not exist then the searching function returns \textsf{low}. From the left we are searching, if it exists, for the smallest index to a value greater than the pivot. If it does not exist, the searching function returns \textsf{high}. Respectively, these task are done by the auxiliary functions \textsf{small-index} and \textsf{large-index} that take as input the given vector interval and the value of the pivot. Observe, that both functions receive a valid vector interval as input. They are designed using structural recursion and, therefore, indexing errors can not occur. In the interest of brevity the development of these functions is not presented. Finally, the value of \textsf{s-index} is tested to see if it is less than or equal to the value of \textsf{l-index} to determine if the size of the vector subinterval they define is less than 2. If so, the vector elements are separated and the position of the pivot in the sorted vector is returned to \textsf{partition!}. Observe, that this value is an element of the given vector interval. If the value returned by \textsf{small-index} is greater than the value returned by \textsf{large-index}, then the values indexed by the two returned values are swapped and the process of separation continues with the vector interval defined by the two returned values. Observe that these returned values form a valid interval into the vector and, therefore, it is safe to recursively process this new vector interval.

There are two primary lessons for an educator to walk away with from this example. The first is that vector intervals are not only useful to eliminate vector indexing errors. They also provide the ability to demonstrate that the ``pieces of the puzzle" are correctly put together. That is, they provide a means that is understandable to beginners to establish that functions are correctly called with arguments (e.g., vector intervals) that clearly communicate the design ideas. The second is that vector intervals allow us to naturally design a mutator (i.e., quicksort in place) without peppering the code with mutations. In fact, mutation is only needed for \textsf{swap}. The reader can compare this development of quicksort with the typical development that appears in any textbook using an imperative language, where mutations are distributed throughout the code they present.

\subsection{Heapsort}
After developing quicksort, students are shown that when the input vector is sorted the abstract running time is $O(n^2)$. That is, it is no better than insertion sorting. Students are then asked if we can do better. Rarely, does a student suggest an answer and the class is introduced to \emph{heaps}. The heap data definition is:
\begin{alltt}
  A heap is either:
    1. empty
    2. a number (the root) and two heaps (left and right), where the number is
       greater than the roots, if any, of the two subheaps
\end{alltt}
Students are then shown that heaps are easily mapped to vectors. A heap whose root is at index \textsf{i} has, if they exist, the root of its left subheap at index \textsf{2i+1} and the root of its right subheap at index \textsf{2i+2}.

The discussion then focuses on how a vector of numbers can be sorted in place using heaps. Clearly, the first step is to mutate the vector into a heap. The details of how this is done and the initial vector interval for it are postponed until the design of the heapifying function is tackled. Once the vector is a heap, students immediately determine that the entire vector must be sorted and, therefore, the correct initial vector interval for the sorting function is [0..(sub1 (vector-length V))]. This design developed by students in class is captured in Figure \ref{hs}.

\begin{figure}[t]
\begin{alltt}
; sort\(!\): VINTV(wholenum wholenum) \(\rightarrow\) (void)
; Purpose: For the given VINTV, sort the vector elements
; Effect: V's elements in the VINTV are rearranged in non-decreasing order
; Assumption: V is a heap and given VINTV is valid for V
(define (sort\(!\) low high)
  (cond [(empty-VINTV? low high) (void)]
        [else (begin
                (swap low high)
                (trickle-down\(!\) low (sub1 high))
                (sort\(!\) low (sub1 high)))]))
\end{alltt}
\caption{The \textsf{sort!} mutator for heapsort.}
\label{sort}
\end{figure}

The design of the sorting mutator starts with the assumptions that the given vector interval is valid for \textsf{V} and that the elements in the given vector interval form a heap. The need to process the entire vector interval suggests using structural recursion. After looking at some examples of sorting using heaps, students realize that if the given vector interval is empty, then \textsf{V} is sorted and the sorting process may stop. On the other hand, students realize that if the given vector interval is not empty then the largest number is the root of the heap. That is, the largest number is the \textsf{low} element of the given vector interval. Therefore, in the sorted vector this number must be indexed by the \textsf{high} element of the vector interval being processed. To achieve this, at every step the root of the heap must be swapped with the \textsf{high} element. Once a swap occurs, \textsf{V[low..high-1]} may no longer be a heap. This requires making it into a heap. Some students correctly observe that \textsf{heapify!} can be used to achieve this. However, this would require traversing the entire vector interval \textsf{[low..high-1]}--a linear operation which would make the algorithm $O(n^2)$ and no better than insertionsort or quicksort. After some class discussion of examples, students realize that all that is needed is to ``trickle down" within \textsf{[low..high-1]} the new element in the root position to its correct position processing at most a number of elements equal to the height of the heap--a $\lg n$ operation making the algorithm $O(n \lg n)$. This trickling down is a new task and, thus, requires an auxiliary function. Once a heap is re-established, sorting continues by processing the rest of the vector interval using structural recursion. This design for sorting is captured in Figure \ref{sort}. Observe that reasoning about vector intervals has allowed for students to clearly understand and design the sorting function for heapsort.

\begin{figure*}[t]
\begin{alltt}
; trickle-down\(!\): VINTV(wholenum wholenum) \(\rightarrow\) (void)
; Purpose: For the given VINTV, re-establish a heap rooted at low
; Effect: Vector elements are moved to have a heap rooted at low
; Assumption: The given VINTV is valid for V
(define (trickle-down\(!\) low high)
    (local [(define rc-index (+ (* 2 low) 2))
            (define lc-index (add1 (* 2 low)))]
      (cond [(> lc-index high) (void)] ; root has no children
            [(> rc-index high)         ; root only has a left child
             (cond [(<= (vector-ref V lc-index) (vector-ref V low)) (void)]
                   [else  (begin
                            (swap low lc-index)
                            (trickle-down\(!\) lc-index high))])]
            [else ; root has two children
              (local [(define mc-index (max-child-index lc-index rc-index))]
                (cond [(>= (vector-ref V low) (vector-ref V mc-index))
                       (void)]
                      [else (begin
                              (swap low mc-index)
                              (trickle-down\(!\) mc-index high))]))])))]
\end{alltt}
\caption{The \textsf{trickle-down!} mutator.}
\label{td}
\end{figure*}

The mutator, \textsf{trickle-down!}, to reheapify after a swap receives as input a valid vector interval and is designed around the number of children\footnote{These are the numbers of the roots of the subheaps.} the root, \textsf{V[low]}, has. Class discussion reveals how to detect the number of children the root has:
\begin{description}
  \item[No children] The index of the left child is greater than \textsf{high} and, therefore, not in the given interval.
  \item[One child] The index of the right child is greater than \textsf{high} and, therefore, not in the given interval.
  \item[Two children] The index of both children are in the given vector interval
\end{description}
The design recipe in \textsf{HtDP} for functions that must detect different conditions states that an answer must be formulated for each condition starting with the non-recursive cases. This is something that students now know very well and it is natural to proceed in this fashion. The simplest case is when the root has no children. In this case, the root is part of a heap and the process stops returning \textsf{(void)}. If there is only one child it must be the left child. If this left child is less than or equal to the root, then there is a heap and, once again, the process stops returning \textsf{(void)}. Otherwise, the root and the left child are swapped. This operation cannot cause an indexing error, because both indices are known to be members of the given vector-interval. The swapping is followed by recursively trickling down the vector interval formed by the index of the left child (just swapped) and \textsf{high}. Students observe that \textsf{[lc-index..high]} is contained in \textsf{[low..high]} and, thus, is a valid vector interval that can safely be recursively processed. If the root has two children, the index of the maximum child is computed. The root is compared with it. If the root is greater than or equal, as before, there is a heap and the mutator stops returning \textsf{(void)}. Otherwise, the root is swapped with the max child and the vector interval formed by the index of the max child and \textsf{high}, \textsf{[mc-index..high]}, is recursively processed. Once again, observe that the swapped indices are members of the given vector interval and, therefore, it is safe to recursively process this new vector interval. This design is captured in Figure \ref{td}.

The mutator \textsf{trickle-down!} does not process the given vector interval using structural recursion. It is clearly generative recursion by generating a new vector interval with each recursive call. This means that a termination argument must be provided according to the design recipe in \textsf{HtDP}. Thanks to the design using vector intervals this task is not onerous. Students observe that with each recursive call the vector interval is getting smaller and that \textsf{high} remains unchanged. This means that the index of the root, \textsf{low}, of what needs to be reheapified is getting larger. Eventually, it will become large enough that it cannot have children within the given vector interval. That is, the indexes of any potential children will be greater than \textsf{high} and the process stops.

\begin{figure*}[t]
\begin{alltt}
; heapify\(!\): VINT(wholenum wholenum) \(\rightarrow\) (void)
; Purpose: For the given VINTV, make the given vector a heap
; Effect: Reorganizes the vector elements to form a heap rooted at low
; Assumption: The given VINTV is valid for V and low > 0
(define (heapify\(!\) low high)
  (cond [(empty-VINTV? low high) (void)]
        [else
          (local [(define parent-index (parent high))]
            (cond [(> (vector-ref V parent-index) (vector-ref V high))
                   (heapify\(!\) low (sub1 high))]
                  [else (begin
                          (swap parent-index high)
                          (trickle-down\(!\) high (sub1 (vector-length V)))
                          (heapify\(!\) low (sub1 high)))]))]))
\end{alltt}
\caption{The Mutator that Transforms a Vector Into a Heap.}
\label{heapify}
\end{figure*}

The only interesting tasks left are to write a mutator that converts an arbitrary vector into a heap and to decide what the initial input vector interval to this function ought to be. The code developed in class is displayed in Figure \ref{heapify}. It is developed using structural recursion processing the given vector interval from right to left. If the vector interval is empty, the process stops as there is nothing to heapify and returns \textsf{(void)}. If it is not empty, the index of the parent of the \textsf{high} element of the vector interval is computed. If the parent is larger than the \textsf{V[high]}, then the parent and the child do not violate the invariant in the data definition of a heap and the rest of the interval is recursively made into a heap. Otherwise, \textsf{V[high]} and its parent are swapped to make the heap invariant hold. This swap may destroy the heap that existed in \textsf{V[high..(sub1 (vector-length V))]} and must be reheapified. This is accomplished by calling \textsf{trickle-down} that was designed above. Once \textsf{V[high..(sub1 (vector-length V))]} is a heap, the rest of the vector interval is recursively processed.

Students commonly want to heapify the vector, \textsf{V}, using  the vector interval \textsf{[0..(sub1 (vector-length V))]}. It is natural for students to gravitate to this choice given that they want to heapify the entire vector. Correcting this requires that they think carefully about their design and about vector intervals. The design of the mutator \textsf{heapify!} revolves around the index of the parent of the \textsf{high} element being in the vector interval. This means that 0 is not part of the vector interval that must be heapified given that it does not have a parent index in \textsf{[0..(sub1 (vector-length V))]}. All other indices have a parent in this vector interval. Therefore, the correct vector interval that must be provided to the initial call to \textsf{heapify!} in Figure \ref{hs} is \textsf{[1..(sub1 (vector-length V))]}.

The important lesson for educators to take away from this example is that an algorithm that many advanced students consider complex is simplified when one reasons about vector intervals. Vector intervals allow beginning programmers to focus on problem solving and reason about vector indexes in their code. They are not left adrift to figure out how to avoid indexing errors when designing using generative recursion. Observe that the focus is not on the required mutations that, once again, are only needed for \textsf{swap}. Equally important, this example demonstrates the usefulness of vector intervals when designing functions, like \textsf{trickle-down!} in Figure \ref{td}, that do not process all the elements of a given vector interval.

\subsection{Radixsort}
After implementing heapsort, students are asked if sorting can be done in linear time. This is a difficult question for them. Some students propose that it can not be done, because any given number in a vector must be compared with at least some other numbers in the vector. They are explained that this line of thought is correct for comparison-based sorting, but that there is another way. Students, to date, can not fantom what this means, but most are intrigued. They are told that instead of comparing numbers with each other, we can process the digits of all numbers from least significant to most significant without comparing them to each other.

To do so, however, it is useful to first design and implement an interface for a data structure that we call a bucket. A bucket has a \textsf{(vectorof X)} and an index for the next available position in the vector (which is equal to the number of elements in the bucket). It is an interface that offers the following services:
\begin{description}
  \item[\textsf{(make-bucket k)}] Creates a bucket that can hold \textsf{k} elements. Initially, all \textsf{k} elements are \textsf{(void)} and the number of elements in the bucket is zero.
  \item[\textsf{(bucket-add! n)}] Puts \textsf{n} in the next spot available in the bucket's vector and increases the number of elements by 1.
  \item[\textsf{(bucket-dump! D i)}] Empties the bucket and puts all its elements in
  vector \textsf{D} starting at index \textsf{i}. Emptying the bucket returns all vector elements to \textsf{(void)} and the number of elements to 0.
  \item[\textsf{(bucket-size)}] Returns the number of elements in the bucket.
  \item[\textsf{(bucket-elems)}] Returns the bucket's vector of elements.
\end{description}
The bucket interface is relatively simple for students to implement. The hardest service to implement is \textsf{bucket-dump!} which is implemented using structural recursion on a vector interval.

Armed with buckets, students are ready to learn about radixsort. To simplify the presentation, students are told to assume the vector contains only nonnegative numbers. The algorithm is outlined as follows:
\begin{enumerate}
  \item Create 10 buckets (one for each possible digit).
  \item Process all the digits of the vector numbers starting with the least significant digits and ending with the most significant digits.
  \item Based on the digit being processed place the number in the corresponding bucket.
  \item After all numbers have been placed in a bucket, dump all the buckets back into the vector starting with bucket 0.
  \item Recursively process the next most significant digits.
  \item When all digits are processed, stop. The vector is sorted.
\end{enumerate}
To further simplify the task, students are told to treat all numbers as having the same number of digits. If a number is not long enough to have a digit, then the digit is 0. For example, the hundreds digit for 9 is 0 (i.e., 009 = 9).

\begin{figure*}[t]
\begin{alltt}
(define (radix-sortin-place\(!\) V)
  (local [ \(\ldots\)
          (define (make-buckets)
            (build-vector 10 (lambda (i) (make-bucket (vector-length V)))))
          (define buckets (make-buckets (vector-length V)))
          (define (sort\(!\)) (helper\(!\) 0))
          (define (helper\(!\) i)
            (cond [(= i sz) (void)]
                  [else (begin
                          (bucketize\(!\) 0 (sub1 (vector-length V)) i)
                          (dump-buckets\(!\) 0 0)
                          (helper\(!\) (add1 i)))]))
          (define (bucketize\(!\) low high i)
            (cond [(empty-VINTV2? low high) (void)]
                  [else
                   (local [(define num (vector-ref V low))
                           (define bucketnum (compute-bucket-number num i))]
                     (begin
                       (bucket-add\(!\) (vector-ref buckets bucketnum) num)
                       (bucketize\(!\) (add1 low) high i)))]))
          (define (dump-buckets\(!\) bnum index)
            (cond [(= bnum 10) (void)]
                  [else
                   (local [(define bucket (vector-ref buckets bnum))
                           (define newindex (+ index  (bucket-size bucket)))]
                     (begin
                       (bucket-dump\(!\) bucket V index)
                       (dump-buckets\(!\) (add1 bnum) newindex)))]))]
    (sort\(!\))))
\end{alltt}
\caption{The outline for Radixsort.}
\label{rs}
\end{figure*}

The basic outline developed in class for radixsort is displayed in Figure \ref{rs}. Contracts, purpose statements, and effect statements from Figure \ref{rs} are omitted to reduce its size. A vector of 10 buckets is created to process the digits. Each bucket can hold, if necessary, all the elements of the vector being sorted. The body of the \textsf{local} calls an auxiliary function, \textsf{sort!}, to sort the vector. In turn, this function calls another auxiliary mutator, \textsf{helper!}, that takes as input the power of 10, \textsf{i}, indicating the position of the digit being processed. At the beginning, this power is 0 for the ones position of all numbers. The mutator \textsf{helper!} is designed using generative recursion on natural numbers. If the given power for 10 is equal to, \textsf{sz}, the length of the largest number, then the process stops returning \textsf{(void)}. Otherwise all the numbers are placed in the buckets, dumped back into the vector, and recursively the next power of 10 is processed. The function will terminate, because with each recursive call its input, \textsf{i}, gets one closer to the length of the largest number in the vector. Eventually, these will be equal and the function halts.

The mutator, \textsf{bucketize!}, puts all the numbers in the vector into the buckets. It takes as input a vector interval for all the numbers to put in buckets and the position of the bigit by which to place numbers in buckets. Initially, the given vector interval spans the entire vector. The function is designed using structural recursion on the given vector interval (from left to right to preserve the relative ordering of the numbers in the vector). If the interval is empty, the function terminates and returns \textsf{(void)} as all vector elements have been placed in a bucket. Otherwise, the number indexed by \textsf{low} is retrieved from the vector, the bucket to place it in is computed, the number is placed in the correct bucket, and the rest of the interval is recursively processed.

Finally, the function, \textsf{dump-buckets!}, to place all the numbers back in the vector takes as input a bucket number and an index into the vector where to start placing back the numbers. The function is designed using generative recursion with natural numbers. If all buckets have been processed, \textsf{(void)} is returned and the mutator halts. Otherwise, it dumps the current bucket into the vector starting at the current position (i.e., \textsf{index}). It then recursively process the next bucket starting at the first index after the last number dumped into the vector. The mutator is guaranteed to terminate, because the process starts with bucket 0 and at each step the bucket number is increased by 1. Eventually, all buckets are dumped and the process stops when the bucket number reaches 10.

The important lesson that educators should walk away from this example is that computations involving vectors that are not structurally recursive, like radixsort, may use structural recursion on vector intervals. Whenever possible structural recursion ought to be used (with beginners), because it is the simplest, it guarantees no indexing errors, and it commonly suffices in terms of efficiency. Generative recursion should be used when significant gains in performance can be achieved. This usually requires profound insights into the problem that is being solved. Vector intervals, nonetheless, have proven to be a useful concept to reason about and design around when using generative recursion with vectors. It is also noteworthy, once again, that the emphasis is on design around vector intervals and not on mutations. All mutations in this case are inside the interface for a bucket. Thus, liberating students to think about the process of radix sorting.

\section{Concluding Remarks}
\label{conclusions}
This article presents a type-driven design-oriented methodology to assist beginners with the development of vector-processing functions using generative recursion. The methodology is based on the concept of a vector interval that has a recursive structure. The examples presented suggest that vector intervals are a useful abstraction when designing vector-processing programs based on generative recursion. They help to determine if functions are (recursively) called with a valid vector interval for a given vector. Calling a function with an invalid vector interval is likely to cause vector indexing errors. As such, vector intervals reduce debugging time and frustration for beginners. They also help in determining if a specific index into a vector is valid. To be valid, an index must be a member of the vector interval that is being processed. Another useful benefit of vector intervals is that termination arguments may be easier to formulate.

Vector intervals have also proven an invaluable tool in exposing students to algorithms that are traditionally considered too advanced for beginners. As far as the author can tell, the overwhelming majority of universities in the world do not expose first-year Computer Science students to heap and radix sorting. In fact, students traditionally struggle with these algorithms in an Algorithms course. Thanks to the development of vector intervals, such algorithms are part of the traditional cannon for our Introduction to Computer Science sequence at Seton Hall University. Vector intervals allow the student and the instructor to focus on the design of solutions to problems and spend much less time on debugging vector indexing errors and on using mutations. The latter is a rather important characteristic, because the focus remains on developing code that others can understand and maintain.

Future work includes formulating accumulative recursion as well as multidimensional-vector processing examples that further demonstrate the usefulness of reasoning about vector intervals to design vector-processing functions. Additionally, students assessments will be pursued to quantify how students feel about the usefulness of designing vector-processing functions using vector intervals. Finally, the work is also being extended to demonstrate the usefulness of vector intervals in a course that focuses on object-oriented design. Vector intervals are a syntax-free notion and as such they are an abstraction that is useful in any programming language that provides support for vector programming.

\section{Acknowledgements}
The motivation to develop the work presented in this article has flowed from the difficulties my beginning students at Seton Hall University have faced. First, they struggled to understand the basics of vector programming. Second, they went on to more advance courses to struggle with complex algorithms involving vector programming. Since the development of this material students have reported less frustration with vectors in introductory Computer Science and have expressed being able to understand complex algorithms in other courses thanks to their exposure to vector intervals. I thank all my Seton Hall University students for inspiring me to find a better way to introduce beginners to vector programming.

\bibliographystyle{eptcs}
\bibliography{vp}
\end{document}